\documentclass[twocolumn]{aastex701mod}

\usepackage[encapsulated]{CJK}  
\newcommand{\cntext}[1]{\begin{CJK}{UTF8}{gbsn}#1\end{CJK}}

\usepackage{colortbl}
\usepackage{xcolor}
\usepackage{gensymb}
\usepackage{amsmath}
\usepackage{tabularx}

\newcommand{\jhu}{William H. Miller III Department of Physics and Astronomy, Johns Hopkins University, Baltimore, MD 21218, USA}
\newcommand{\jhuapl}{Johns Hopkins University Applied Physics Laboratory, 11100 Johns Hopkins Road, Laurel, MD 20723}

\newcommand{\goddard}{NASA Goddard Space Flight Center, 8800 Greenbelt Road, Greenbelt, MD 20771, USA}

\newcommand{\oslo}{Institute of Theoretical Astrophysics, University of Oslo, P.O. Box 1029 Blindern, N-0315 Oslo, Norway}
\newcommand{\cwru}{Department of Physics, Case Western Reserve University, 10900 Euclid Avenue, Cleveland, OH, 44106 USA}
\newcommand{\cfa}{Center for Astrophysics, Harvard \& Smithsonian, 60 Garden Street, Cambridge, MA 02138, USA}

\newcommand{\ucsc}{Departamento de Ingenier\'{i}a El\'{e}ctrica, Universidad Cat\'{o}lica de la Sant\'{i}sima Concepci\'{o}n, Alonso de Ribera 2850, Concepci\'{o}n, Chile}

\newcommand{\uchicago}{Department of Astronomy and Astrophysics, University of Chicago, 5640 South Ellis Avenue, Chicago, IL 60637, USA}
\newcommand{\kicp}{Kavli Institute for Cosmological Physics, University of Chicago, 5640 South Ellis Avenue, Chicago, IL 60637, USA}
\newcommand{\chicagophysics}{Department of Physics, University of Chicago, 5640 South Ellis Avenue, Chicago, IL, 60637, USA}

\newcommand{\umbc}{The University of Maryland, Baltimore County. Baltimore, Maryland, USA}

\submitjournal{ApJ}

\shorttitle{CLASS 4-Year Circular Polarization}
\shortauthors{Essinger-Hileman et al.}

\begin{document}

\title{Improved Constraints on Cosmic Microwave Background Circular Polarization with CLASS}

\correspondingauthor{Thomas Essinger-Hileman}
\email{tom.essinger-hileman@nasa.gov}
\author[0000-0002-4782-3851]{Thomas Essinger-Hileman}\affiliation{\goddard}
\author[0009-0001-1748-7877]{Caleigh Ryan}\affiliation{\jhu}
\author[0000-0002-4820-1122]{Yunyang Li (\cntext{李云炀}\!\!)}\affiliation{\kicp}
\author[0000-0002-4436-4215]{Matthew A. Petroff}\affiliation{\cfa}
\author[0000-0002-8412-630X]{John~W. Appel}\affiliation{\jhu}
\author[0000-0001-8839-7206]{Charles L. Bennett}\affiliation{\jhu}
\author[0000-0001-8468-9391]{Ricardo Bustos}\affiliation{\ucsc}
\author[0000-0001-8144-556X]{Carol Yan Yan Chan}\affiliation{\jhu}
\author[0000-0003-0016-0533]{David T.~Chuss}\affiliation{Department of Physics, Villanova University, 800 Lancaster Avenue, Villanova, PA 19085, USA}
\author[0000-0002-7271-0525]{Joseph~Cleary}\affiliation{\jhu}
\author[0000-0002-1708-5464]{Sumit Dahal}\affiliation{\goddard}\affiliation{\jhu}\affiliation{\jhuapl}
\author[0000-0003-3853-8757]{Rahul Datta}\affiliation{\uchicago}\affiliation{\jhu}
\author[0000-0002-0552-3754]{Jullianna Denes~Couto}\affiliation{\jhu}
\author[0000-0002-3592-5703]{Kevin~L. Denis}\affiliation{\goddard}
\author[0000-0001-6976-180X]{Joseph R. Eimer}\affiliation{\jhu}
\author[0000-0002-4421-0267]{Joseph E. Golec}\affil{Department of Astronomy, University of Massachusetts Amherst, 710 N. Pleasant Street, Amherst, MA 01003, USA}
\author[0000-0001-9238-4918]{Kyle R. Helson}\affiliation{\goddard}\affiliation{\umbc}
\author[0000-0001-7466-0317]{Jeffrey Iuliano}\affiliation{\jhu}
\author[0000-0003-4496-6520]{Tobias~A. Marriage}\affiliation{\jhu}
\author[]{Jeffrey~J. McMahon}\affiliation{\kicp}\affiliation{\uchicago}\affiliation{\chicagophysics}
\author[0000-0002-0024-2662]{Ivan Padilla}\affiliation{\cwru}
\author[0000-0002-1371-5334]{Carolina Morales Perez}\affiliation{\jhu}
\author[0000-0003-4189-0700]{Karwan Rostem}\affiliation{\goddard}
\author[0000-0001-7458-6946]{Rui Shi (\cntext{时瑞}\!\!)}\affiliation{\jhu}
\author[0000-0003-3487-2811]{Deniz A. N. Valle}\affiliation{\jhu}
\author[0000-0002-5437-6121]{Duncan J. Watts}\affiliation{\oslo}
\author[0000-0003-3017-3474]{Janet L. Weiland}\affiliation{\jhu}
\author[0000-0002-7567-4451]{Edward J. Wollack}\affiliation{\goddard}






\begin{abstract}
We present improved constraints on circular polarization in the cosmic microwave background (CMB) from the Cosmology Large Angular Scale Surveyor (CLASS) experiment using data at 90, 150, and 220~GHz to complement previous results based on 40~GHz data alone. We extend constraints on circular polarization from the atmosphere to all CLASS frequencies and compare with a model of the atmospheric circular polarization spectrum. The inclusion of data at all CLASS frequencies improves upper limits on circular polarization, Stokes $V$, in the CMB by a factor of $\sim 5$ at the largest angular scales ($10 < \ell < 30$) and extends the constraints to smaller angular scales, $\ell = 700$. In combination with \textit{Planck} temperature data, we provide the first limits on the temperature-circular-polarization ($TV$) power spectrum to date. Both circular polarization ($VV$) and $TV$ power spectra are consistent with zero, and CLASS places upper limits of $5.4 \times 10^{-3}$~$\mu$K$_{\text{CMB}}^2$ on $VV$ at large angular scale ($10 < \ell < 30$) and $5.8 \times 10^{-1}$~$\mu$K$_{\text{CMB}}^2$ on $TV$ at 95\% confidence at $30 < \ell < 50$.

\end{abstract}
\keywords{Cosmic microwave background radiation, Early universe, Observational cosmology, Astronomical instrumentation, Polarimeters, Sky surveys, Reionization, Big Bang theory}


\section{Introduction} \label{sec:intro}
\setcounter{footnote}{0}

Measurements of the cosmic microwave background (CMB) temperature and linear polarization anisotropies provide stringent constraints on the state of the early universe. The \textit{Planck} and Wilkinson Microwave Anisotropy Probe (WMAP) satellite missions have measured the temperature anisotropy to cosmic variance limits on angular scales $\gtrsim 0.1^\circ$ ($\ell \lesssim 1600)$, and place the tightest constraints to date on the even-parity \textit{E} mode linear polarization at angular scales $\gtrsim 10^\circ$ ($\ell \lesssim 20)$~\citep[e.g.,][]{hinshaw13, planck18VI}. Ground-based and balloon-borne microwave telescopes provide complementary data that constrain small-scale temperature anisotropy for $\ell \lesssim$ 10\,000, $E$ mode polarization at intermediate scales $\ell \approx$~20--2000, and the odd-parity $B$ mode polarization from gravitational lensing of $E$ modes~\citep[e.g.,][]{spider21, pb20emode, Louis2025, Camphuis2025}. Together with measurements of the expansion history and large scale structure~\citep[e.g.,][]{spt19clustercosmology, kids20shear, hsc23shear,eboss21baocosmology,des22baocosmology, pantheon22cosmology, desi_cosmology_2025}, these observations provide strong support for a universe dominated by dark energy (modeled as a cosmological constant, $\Lambda$) and cold dark matter (CDM) and place constraints on the parameters of this concordance $\Lambda$CDM cosmological model at the $\sim1$\% level. Searches for $B$-modes from a primordial gravitational wave background from inflation are ongoing, with the most stringent limits coming from a combination of ground and satellite data~\citep[][]{BK21}. 

The primary CMB is not expected to be circularly polarized within the $\Lambda$CDM paradigm, and any circular polarization arising from inflation is strongly suppressed~\citep{alexander_generation_2009}. Consequently, searches for CMB circular polarization are highly sensitive to physics beyond the Standard Model as well as astrophysical processes. 

Primordial and beyond-standard-model (BSM) physics generate circular polarization through exotic physics. Possible mechanisms for primordial CMB circular polarization include scattering processes in the presence of a primordial magnetic field, Lorentz invariance violations, and non-commutative field theories~\citep{King2016_CMB_Circular_Pol}. Certain BSM sources, such as axions coupling to CMB photons~\citep{alexander_physics_2020} naturally produce correlated CMB temperature ($T$) and circular polarization ($V$). As the CMB temperature has been well characterized by \textit{Planck} and WMAP at the angular scales probed by the Cosmology Large Angular Scale Surveyor (CLASS) ($\ell < 700)$, these sources of $V$ polarization may be best constrained through the $TV$ cross-spectrum. The presence or absence of $TV$ correlation would distinguish between possible production mechanisms.

Secondary cosmological and propagation effects induce circular polarization at later times as CMB photons travel through the universe. Birefringence can arise from the anisotropy of the CMB surface of last scattering, which alters the polarization state of the light, although this signal is expected to be much smaller than later astrophysical sources~\citep{inomata_circular_2019}. At later times, weak gravitational lensing by large scale structure can induce circular polarization due to the optical Magnus effect~\citep{Yusuke2026_Optical_Magnus_Vpol}.

Astrophysical and Galactic foregrounds produce circular polarization at microwave wavelengths, either intrinsically or by altering existing radiation. Faraday conversion creates a phase offset between the orthogonal components of linearly polarized radiation in a birefringent medium, resulting in circularly polarized light. This is anticipated to occur when linearly polarized CMB radiation interacts with the magnetized plasma around the first stars and in galaxy clusters. Finally, Galactic synchrotron emission is the strongest known source of circularly polarized microwave radiation~\citep{Legg_Westfold_1968_Elliptical_Synchrotron}.

Celestial circular polarization at microwave frequencies has been constrained previously by \citet{Lubin1983_CMB_pol}, MIPOL~\citep{MIPOL_V_Limit2013}, and SPIDER~\citep{SPIDER_V_Limit_2017}. MIPOL employed coherent detector technology that gave intrinsic sensitivity to circular polarization at 33~GHz. SPIDER exploited non-ideal behavior in their half-wave plate polarization modulator, which is designed to rotate linear polarization, to place constraints on circular polarization in two bands at 90 and 150~GHz. 

The CLASS experiment is an array of microwave telescopes operating at high altitude in the Atacama desert of Chile that has observed in four bands centered near 40, 90, 150, and 220~GHz. While the primary science goals of CLASS are to improve constraints on $E$ mode polarization from reionization and search for inflationary $B$ modes, CLASS has unique sensitivity to circular polarization through its use of variable-delay polarization modulators (VPMs). The CLASS VPMs modulate the sky signal in both linear and circular polarization at a frequency $\gtrsim 10$~Hz to place the polarized signal band above the frequencies associated with drifts in receiver temperatures or unpolarized atmospheric emission~\citep{chus12vpm, harrington18spie, harrington21, cleary22}.

The atmosphere is circularly polarized through Zeeman splitting of oxygen emission lines in the presence of the Earth's magnetic field, presenting a possible contaminant for searches for celestial circular polarization~\citep{Lenoir_oxygen_spectrum, Rosenkranz_1988_oxygen_pol, Hanany_atmosphere_pol_2003}. Fortunately, due to the uniform mixing of atmospheric oxygen, the atmospheric circular polarization signal is a smooth dipole pattern, the direction of which is set by the direction of the Earth's magnetic field-axis. CLASS has previously reported measurements of this atmospheric circular polarization \citep{petroff20} and has placed the most stringent constraints on microwave Celestial circular polarization to date \citep{Padilla_2020_CLASS_Circular, Eimer_2023_CLASS_Results} at 40~GHz. \textsc{Polarbear} has made measurements of the circular polarization of the atmosphere near the CLASS site by exploiting non-ideal behavior in an HWP with an effective band estimated to be at 121.8~GHz~\citep{polarbear_atm_Vpol_2025}.

This paper expands the previous CLASS constraints to include data from the remaining telescopes at 90, 150, and 220~GHz. Furthermore, in combination with \textit{Planck} data, we place the first constraints to date on CMB temperature-circular-polarization ($TV$) cross-correlation.

\begin{figure*}
    \centering
    \includegraphics[width=\textwidth]{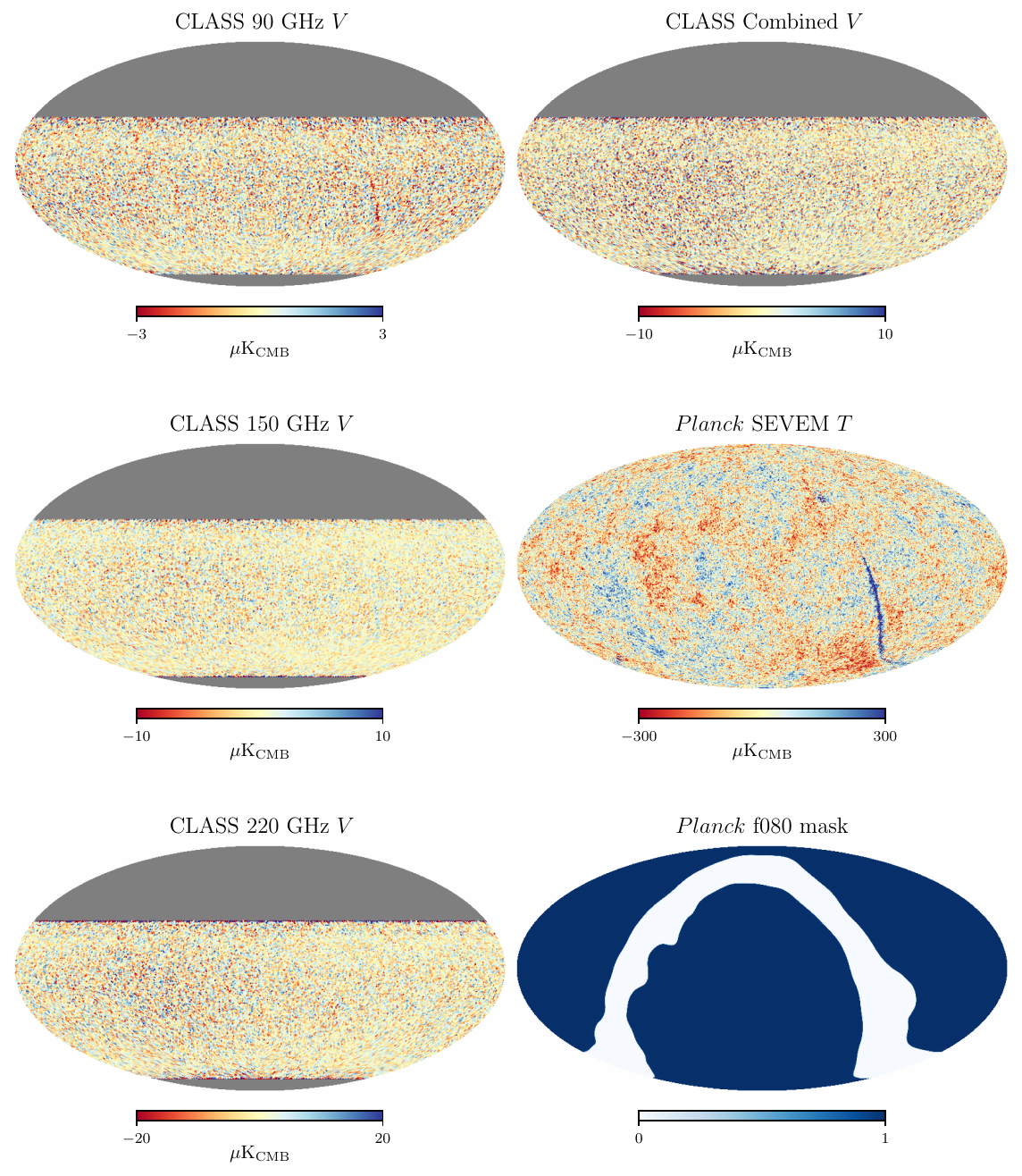}
    \caption{Cosmology maps used in this analysis. All maps are in Celestial coordinates. CLASS maps cover declinations observed from the Chilean site, from $-74^\circ$ to $28^\circ$. The CLASS combined map is a minimum variance co-add of the CLASS maps at 90, 150, and 220~GHz. The non-uniform noise in the CLASS maps is due to uneven coverage in the survey. The \textit{Planck} map (middle right) is the PR4 CMB map using the SEVEM pipeline. The \textit{Planck} mask used for this analysis excludes the brightest 20\% of the sky near the Galactic plane.
    \label{fig:v_celestial_maps}}
\end{figure*}

\section{Data Processing and Mapmaking}
The CLASS telescopes operate from a high-altitude site at $-23^\circ$ latitude and perform continuous $720^\circ$ scans at a constant elevation of $45^\circ$, enabling coverage of approximately $70\%$ of the sky each day. The telescopes also step through daily boresight rotations from $-45^\circ$ to $+45^\circ$ in $15^\circ$ increments to improve linear polarization angle coverage.

The data processing and mapmaking follow the procedure of previous CLASS linear polarization maps \citep{Eimer_2023_CLASS_Results} and are detailed in \cite{Li23} and \cite{li25}. 
Here, we summarize the key steps and highlight differences from the previous treatment.

\subsection{Data Selection}
The three frequency bands of the data were selected differently depending on the survey start time. 
For the 90~GHz channel, we used data from 2018-06-21 through 2024-05-14. 
For the dichroic receiver operating at 150/220~GHz \citep{dahal20HF}, we used the data from 2019-10-17 to 2023-08-18. 
Unlike \cite{li25}, we included the data when a plastic closeout was installed on the telescope forebaffle extension for the cosmology maps.
Although these data are subject to spurious linear polarization signals when the closeout is deformed by the wind, these events have minimal impact on circular polarization data with sufficient filtering \citep{Li23}.
As in \cite{li25}, the 10-minute data packages (DataPkgs) were selected based on weather conditions and instrument status.
We restricted our selection to nighttime data only, when the Sun is below the horizon. 
The selected DataPkgs from each continuous night of observation were concatenated to form ``spans'', which typically comprise $\sim$ 80 DataPkgs. 
Further selections were made based on the Moon avoidance, precipitable water vapor (PWV) magnitude, detector readout conditions, cryostat health, and detector responses (as monitored by the amplitude of the VPM-synchronous emission).
To probe anomalies in the \emph{polarization} data, we also employed quality cuts after the demodulation process (see Section \ref{sec:demod_and_data_proc}, below) as described in \citet{Li_thesis}.
The total amount of data used for the circular polarization maps is 323, 231, and 82 detector-years (det-yr) for 90, 150, and 220~GHz, respectively.

\subsection{Demodulation and Data Processing}\label{sec:demod_and_data_proc}
CLASS uses polarization-sensitive, transition-edge-sensor (TES) bolometers~\citep{rostem12spie, appel14spie, nunez23asc} with fast, front-end VPMs, which consist of a polarizing wire grid in front of a movable mirror that convert between linear and circular polarization states.
The modulation as a function of the VPM grid-mirror distance is described by the VPM transfer function that depends on the reflection, transmission, and alignment of the mirror and the grid, and the (bandpass-integrated) spectrum of the linear and circular polarization sources.  
As discussed in Section \ref{sec:atm signal}, the dominant circular polarization signal at 40, 90, and 150~GHz originates from the atmosphere \citep{petroff20} due to the Zeeman splitting of the oxygen molecules in the Earth's magnetic field, which has a spectrum with zero-crossings at the 90 and 150~GHz passbands \citep{dahal22}. 
Due to uncertainties in the atmospheric spectrum modeling, we did not use this spectrum for our modulation transfer function but used a generic blackbody instead.
A mis-specification of the transfer function could leak linear into circular polarization, which would degrade our upper limit but should not bias the measurement of the circular polarization signal lower. We find the expected variation in amplitude from our transfer function choice to be at the percent-level. We attributed this as a systematic uncertainty and included it in the reported error bars. The transfer function bias calculation is discussed in Section \ref{sec:transfer_func_bias}.

Demodulation shifts the circularly polarized signal from high frequencies ($\sim 20$~Hz) to low frequencies that are set by the scanning speed and beam size. Two sets of data are created from the demodulated timestreams. The first is the ``full-resolution" dataset, where the demodulated data are low-pass filtered to avoid aliasing of power into the signal bandwidth when downsampled. They are cut off at frequencies chosen to accommodate Nyquist sampling of the beam given the scanning speed, which are 2.7, 4.4, and 6.3~Hz for 90, 150 and 220~GHz, respectively.

The second dataset includes a more aggressive low-pass filter of 1.04~Hz for more efficient simulation in the downstream processing and the analysis of the atmospheric signals. 
We denote this the ``low-pass filtered" dataset, to distinguish it from the full-resolution data. 

\subsubsection{Transfer Function Bias} \label{sec:transfer_func_bias}
As mentioned above, a band-integrated assumed sky spectrum is used for demodulating the linear and circular polarization components. For this analysis, the circular polarization spectrum is assumed to be a generic blackbody, rather than the simulated atmospheric spectrum from \citet{petroff20} to avoid biasing from a combination of uncertainties in the model spectrum and in the measured CLASS bandpasses. We used simulations to bound the systematic bias on the circular polarization amplitude from using an incorrect VPM transfer function, either from uncertain assumptions about the circular polarization spectrum or inaccurate modeling of the VPM and the detector bandpasses. Specifically, we selected two circular polarization spectra and generated simulated modulated time-ordered data (TOD) assuming bandpasses sampled from the measured uncertainties on the band center and bandwidth per band; these simulations were then demodulated with the transfer function used for the data. The two assumed spectra are a frequency-independent spectrum that is flat in intensity and the atmospheric model spectrum described in Section \ref{sec:atm signal}, from \citet{petroff20}. These two extreme cases should encapsulate the variation introduced from the choice of circular polarization spectrum. The resultant variation is 2.8\%, 1.2\%, 5.8\%, and 0.37\% for 40, 90, 150, and 220 GHz, respectively. The range in these percentages is attributed to the behavior of the transfer function, particularly near the edges of the VPM throw, when the assumed atmospheric spectrum is complicated. 

\begin{figure*}
    \centering
\includegraphics[width=1.9\columnwidth]{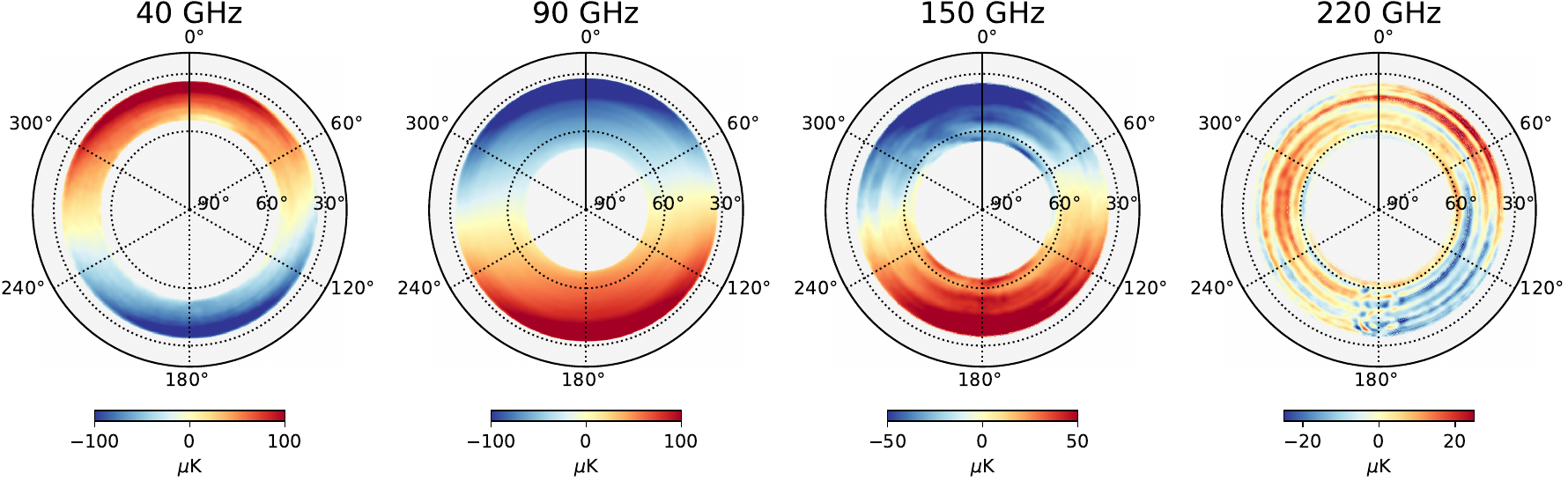}
    \caption{Circular polarization maps at the four CLASS frequency bands in horizontal coordinates smoothed to 2$^\circ$ resolution to emphasize the dipole signal. Plotting the maps in this frame bins the Zeeman splitting signal in azimuth to make a clear dipole for 40, 90, and 150 GHz. The signal is not detected at 220 GHz because the map is dominated by noise and residual systematics. Equation \ref{eq: dipole} is fit to each of these maps and the result is reported in Table \ref{tab:dipole_fit_data}.
    \label{fig:vatm_maps}}
\end{figure*}
\subsection{Mapmaking}\label{sec:mapmaking}

Two sets of maps are created, one for cosmological studies and another for atmospheric studies. The first set of maps are used in Section \ref{sec:vv_spectra} for the celestial circular polarization constraints, and are referred to as the ``cosmology maps.'' They are made from the full-resolution dataset using the same filtering procedure described in \citet{li25}\footnote{A set of cosmology maps is also created using the low-pass filtered dataset to make the noise simulations step more efficient, as described in Section \ref{sec:sims}. However, the cosmology maps used for the celestial analysis are full-resolution.}. Two series of harmonic filters are applied, one for removing the first 12 modes of the $8\pi$ periodicity harmonics associated with the azimuth velocity, and a second series of the first 12 modes of the $2\pi$ periodicity harmonics to remove signals in the azimuth domain. The telescope scanning pattern of alternating 720$^\circ$ azimuth rotations in each direction sets the period for the 8$\pi$ harmonics. Each 720$^\circ$ rotation is referred to as a sweep, while all of the sweeps in a single day form a span. The 8$\pi$ filters are fit to sections of 8 sweeps at a time, while the first six modes of the 2$\pi$ set are fit to 30 sweeps and the last six modes are fit over the entire span. Additional low-order polynomials are fit to the entire span of data to remove any baseline drift and improve the signal stability. 

For the ``atmospheric map" set, no harmonic filtering is applied to the data in order to preserve the atmospheric signal, which is otherwise effectively removed. They are created using the low-pass filtered dataset since the signal of interest is on very large angular scales, and mapped in horizontal coordinates (azimuth and elevation) to make the expected atmospheric signal clearer. We used a filter-and-bin mapmaking where the filtered TOD are weighted by the white noise level in each 10-minute chunk of the data.\footnote{We find convergence issues in the standard maximum-likelihood pipeline with template iteration \citep{Li23}, possibly due to poor cross-linking of the scan pattern in these coordinates.} These maps are used in Section \ref{sec:atm signal} to characterize the circularly polarized atmospheric signal. 

Data taken with a plastic closeout installed on the forebaffle are excluded from the atmospheric maps, but left in the cosmology maps. Deformation of the plastic closeout while facing the wind is only expected to produce linear polarization, but there may be some linear to circular polarization leakage of this signal. This would produce a systematic error in the measured dipole since the wind at the CLASS site primarily comes from one direction, making it difficult to fully separate from the atmospheric dipole. We tested this at 90 GHz, and found that the inclusion of closeout-on data produced a $\sim6 ^{\circ}$ clockwise shift in the measured dipole direction. Excluding the closeout-on data removes this contaminant from the atmospheric maps.
The filtering used for the cosmology maps effectively removes the wind signal. This allows data with the plastic closeout installed to be used for that portion of the analysis, increasing the map depth achieved.

Total circular polarization (Stokes $V$) maps derived from all data for CLASS 90, 150, and 220~GHz, as well as the combined map across the three bands, are shown in Figure~\ref{fig:v_celestial_maps}, along with the \textit{Planck} SEVEM $T$ map~\citep{planck18IV} used for the $TV$ cross-spectrum analysis and the associated mask removing 20\% of the sky near the Galactic plane. All maps are in thermodynamic CMB temperature units.

\subsection{Noise simulations}
\label{sec:sims}
Noise simulations are used extensively in this work for null tests and uncertainty estimates. For the cosmology maps, the simulations were created following the same procedure as previous results \citep{Li23,li25}, while demodulated TOD simulations were generated from the noise model estimated at the mapmaking step. 
These simulated timestreams were then mapped in the same way as the data, as described in Section~\ref{sec:mapmaking}.
However, this approach is less efficient for the full-resolution maps at 90, 150, and 220~GHz. 
Leveraging the fact that the high-frequency component of the demodulated data and the high-$\ell$ component of the maps are close to white noise, we estimated the inhomogeneous white noise level in the full-resolution maps and sampled the noise directly at the map level. 
These white noise simulations are stitched with the maps from the low-pass filtered simulations in the harmonic space to ensure a continuous transition of the noise spectra from low- to high-$\ell$ \citep{Li_thesis}.
It should be noted that this procedure makes a critical assumption that the full-resolution maps have the same low-$\ell$ properties as those of the low-pass filtered maps. 
This is not guaranteed to be valid since the noise models used for the maximum-likelihood mapmaking \citep{dunner13,Li23} are estimated from different frequency ranges. 
However, in practice, we find this makes little difference to the full-resolution maps, and the simulations generated from this prescription match well with the data for all the consistency tests.

For the atmospheric maps, a ``sign flip'' method was used to better capture noise in the large-scale dipole pattern. For these, half of the data, in units of sweeps, is randomly assigned a coefficient of minus one and then combined with the other half of the data in mapmaking. This effectively removes the constant sky signal, and leaves the noise and any systematic variations that are not distributed evenly across the data. Many realizations of these sign-flip noise maps were created to obtain uncertainties on the dipole parameters fit to the data in Section \ref{sec:atm signal}.

\subsection{Null Tests}
\begin{deluxetable}{c|c|c|c}
\setlength{\arrayrulewidth}{.09em}
\tablecaption{\label{tab:null-pte}
Null test PTE table. 
The PTE values are tabulated for every null test for each frequency band. Values less (greater) than 0.5 mean that the simulation shows less (more) scatter than the data. A subscript third significant digit is included in the cell otherwise rounding to 1. Values falling outside of the range 0.05 to 0.95 (the 2$\sigma$ quantiles) are marked in bold. 
}
\tablehead{\colhead{Split} & \colhead{$90\,\mathrm{GHz}$} & \colhead{$150\,\mathrm{GHz}$} & \colhead{$220\,\mathrm{GHz}$}} 
\startdata
top/bot & 0.15  & \textbf{0.04} & 0.89 \\
left/right & 0.07 & 0.81 & 0.64 \\
radial & 0.19 & 0.95 & 0.18\\
horizontal & \textbf{0.04} & 0.60 & 0.85 \\
vertical & \textbf{0.99$\mathbf{_9}$} & 0.57 & 0.45\\
quadrupole & 0.59 & 0.73 & 0.61 \\
MUX halves & 0.79 & 0.77 & 0.86 \\
MUX parity & \textbf{0.02} & 0.70 & 0.21 \\
VPM syn. & 0.30 & 0.10 & 0.24 \\
bs in/out & 0.42 & 0.48 & 0.21 \\
bs pos/neg & 0.28 & 0.69 & 0.52 \\
az velocity & 0.41 & 0.08 & 0.17 \\
6h in/out & 0.53 & 0.51 & 0.78 \\
pre-mid/post-mid & 0.71 & 0.56 & 0.40 \\
half1/half2 & 0.74 & \textbf{0.98} & 0.90 \\
moon up/down & 0.72 & 0.71 & 0.48
\enddata
\end{deluxetable}
To test for systematics in the data, null tests were performed, following the procedure described in \citet{li25}. For each null test, the data were divided in half either by detector location on the focal plane or temporally. Thus divided, we call each half of the data a "data split". For each data split, four base maps were constructed with equal numbers of spans selected to evenly distribute different instrument configurations and observation conditions. Next, null maps were created by pair-differencing the two splits for each of the four base maps. Six cross-spectra between all pairs of null maps were calculated and averaged to obtain the final null spectrum. This process was repeated for 50 noise-only simulations per data split, with the cross spectra of all simulations being calculated to result in 2500 simulated null spectra per null test to compare with the data null spectrum (50 spectra for data split A times 50 spectra for data split B). From this dataset, the probability-to-exceed (PTE) is calculated and reported in Table \ref{tab:null-pte}.  

For individual PTE values, values falling outside of the range 0.05 to 0.95 (the 2$\sigma$ quantiles) are marked in bold. However, since there are 3$\times$16 independent tests performed in total, our criterion for PTE, accounting for the look-elsewhere effect, is between 0.001 and 0.999. No test in Table \ref{tab:null-pte} is found to have an extreme PTE value. We also performed a Kolmogorov-Smirnov (KS) test to confirm that the PTE distributions are consistent with uniform, with KS PTEs of 0.45, 0.74, and 0.68 for 90, 150, and 220 GHz, respectively.

\section{Atmospheric Circular Polarization}\label{sec:atm signal}

Zeeman splitting of atmospheric molecular oxygen is the dominant signal seen in the CLASS atmospheric circular polarization maps, which are plotted in horizontal (azimuth, elevation) coordinates (Figure \ref{fig:vatm_maps}). The direction and amplitude of the polarized signal depend on Earth’s magnetic field, resulting in a smooth dipole-like pattern on the sky\footnote{The signal is ``dipole-like" because it does not follow the spherical harmonic definition of a dipole, due to it scaling with airmass as the tangent of the zenith angle. However, at a constant elevation it is a sinusoidal dipole. The signal will be referred to as a dipole in the rest of the section}. Measurements of the atmospheric circular polarization signal have been previously reported by CLASS at 40~GHz and by \textsc{Polarbear} in the 150~GHz band (effective frequency for circular polarization at 121.8~GHz)~\citep{petroff20, polarbear_atm_Vpol_2025}. Here we report further measurements by CLASS of the atmospheric circular polarization at 90, 150, and 220~GHz. 

CLASS circular polarization maps constructed in horizontal coordinates with minimal filtering effectively bin the dipole signal in azimuth over the full observing period. The maps shown in Figure~\ref{fig:vatm_maps} are fit with the following equation:
\begin{equation}\label{eq: dipole}
    V = a\cdot \text{tan}(b\cdot \phi)\cdot \text{cos}(\psi - c) + d,
\end{equation}

\noindent where $\phi$ is the zenith angle, $\psi$ is the azimuth angle, and $a$, $b$, $c$, and $d$ are the free parameters to be fit. Parameter $a$ is the dipole amplitude, $b$ sets the zenith-angle scaling, $c$ is the dipole direction, with positive (negative) values corresponding to clockwise (counterclockwise) rotation from $0^\circ$ azimuth, and $d$ accounts for uniform (monopole) offset in the circular polarization signal. The azimuth angle of Earth's magnetic field at the CLASS site shifted from $-5.8^\circ$ to $-7.5^\circ$ between 2016 and 2024, calculated using the 13th generation International Geomagnetic Reference Field \citep{alken_international_2021}. We expect the dipole from our maps to point toward $-7.2^\circ$ for 90, 150, and 220 GHz and toward $-6.5^\circ$ for 40 GHz, corresponding to the magnetic field direction near the midpoint of each observing period.  

\begin{figure*}
    \centering
    \includegraphics[width=0.7\textwidth]{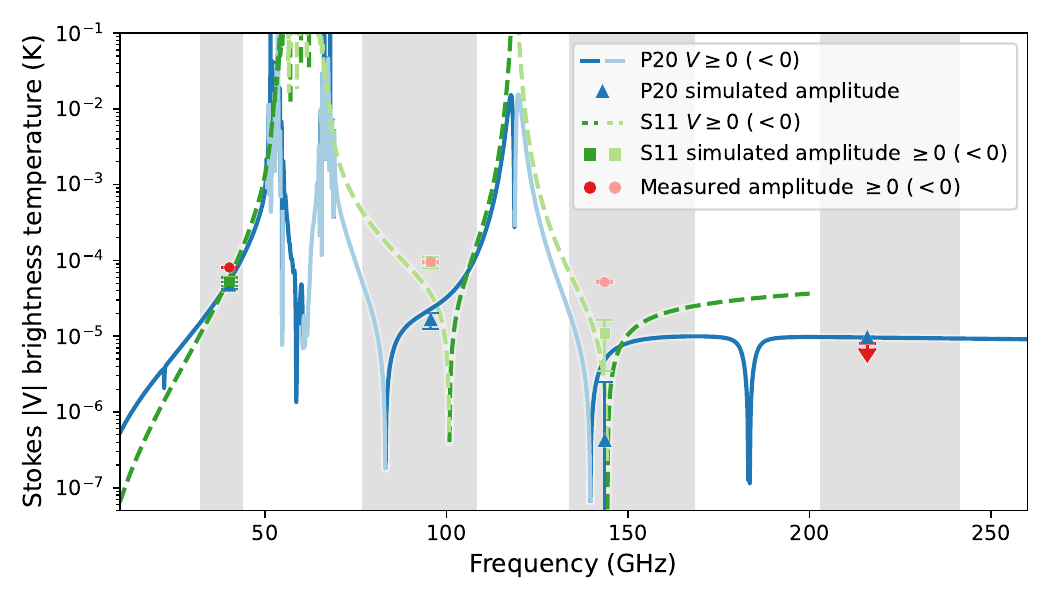}
    \caption{The simulated and measured circular polarization spectrum at the four CLASS frequency bands. The blue and green curves show the modeled atmospheric spectra of \cite{petroff20} (P20) and \cite{spinelli_2011} (S11), respectively. Gray shaded regions indicate the four CLASS bandpasses. Simulated dipole amplitudes from P20 and S11 are shown as blue triangles and green squares, respectively, while red circles show the best-fit amplitudes from the data using Equation \ref{eq: dipole}. For each set of points, darker shades denote positive model or data values and lighter shades denote negative values. The P20 model disagrees significantly with the measured dipole amplitudes at 90 GHz and 150 GHz, whereas the S11 model shows improved agreement at 150 GHz and exact agreement at 90 GHz. Error bars on the red circles give the standard deviation of the dipole amplitude from 200 noise simulations; error bars on the simulated points include uncertainties from the transfer function bias and the CLASS bandpasses. Model uncertainties in the theory that defines the model are the suspected cause of discrepancies, especially far from the line centers.  
    \label{fig:vatm_spec}}
\end{figure*}

The best-fit parameters are reported in Table \ref{tab:dipole_fit_data} and the corresponding dipole amplitudes are shown as the red and pink circular points in Figure \ref{fig:vatm_spec}. Residual systematics in the map, resulting in the elevation stripes visible in Figure \ref{fig:vatm_maps}, caused instability in the fit. We fixed $b$ to its nominal value of 1, resulting in improved stability. The associated errors for each reported parameter are calculated using the 200 sign-flip noise maps, described in Section \ref{sec:sims}. We repeated the fitting for each noise map with the best-fit dipole injected into the map. The standard deviations of the resulting best-fit parameter distributions are reported as the uncertainties in Table \ref{tab:dipole_fit_data} and as the error bars on the red and pink circular points in Figure \ref{fig:vatm_spec}. The dipole signal is clearly detected at 40, 90, and 150 GHz, but 220 GHz appears dominated by systematics. While the signal is not detected at 220 GHz, the data put a meaningful upper limit on the dipole amplitude that is lower than the \citet{petroff20} model prediction.

As discussed in Section \ref{sec:mapmaking}, we do not apply filtering to the atmospheric maps to preserve the dipole, meaning systematics are also left in the maps. To produce a stable fit for 220 GHz, we additionally fix the dipole direction to its expected value of $-7.2^\circ$. 

To quantify how well the dipole model describes each map, we calculated the coefficient of determination ($R^2$) for each fit. Because $R^2$ is the most straightforward to interpret for a linear model, we repeated the fit with parameters $b$ and $c$ fixed to their theoretical values. For $b$ this is $1$ for all bands, while $c$ is the direction near the center of each observing period. We compared the residual between the three-parameter fit and the two-parameter fit and found little impact on the inferred amplitudes. $R^2$ quantifies how much of the variance in the data is explained by the model. An $R^2$ of $1$ means that all of the variance in the data is explained by the model, while $0$ corresponds to a fit in which none of the variance is explained. We calculate $R^2$ values of 0.95, 0.99, 0.97, and 0.24 for 40, 90, 150, and 220 GHz, respectively. The low $R^2$ value at 220 GHz indicates that we do not detect the Zeeman splitting signal in that band, so we report an upper limit on the amplitude in Table \ref{tab:dipole_fit_data} and in Figure \ref{fig:vatm_spec}.

\begin{table*}[t]
    \footnotesize
    \caption{Best fit parameters from Equation \ref{eq: dipole} for Stokes $V$ maps in horizontal coordinates. Parameter $b$ is held fixed to improve the stability of the fitting. For 220 GHz, $c$ is also held fixed, and the values for $a$ and $d$ are the 95\% CL. Uncertainties are estimated using 200 noise simulations.} 
    \begin{tabular}{c|c|c|c|c}
        \hline\hline
         Band & a ($K$) & b & c ($^\circ$) & d ($K$) \\
         \hline
         
         40 GHz & $8.082 \times 10^{-5} \pm 2.026\times10^{-6}$ & $-$ & $-7.697 \pm 0.199$ & \phs $1.905 \times 10^{-6} \pm 1.402 \times 10^{-7}$ \\
         
         90 GHz & $-9.555 \times 10^{-5} \pm 2.630\times10^{-6}$ &  $-$ & $-6.697 \pm 0.185$ & $-1.106 \times10^{-6} \pm 5.394 \times 10^{-8}$ \\ 
         
         150 GHz & $-5.166\times10^{-5} \pm 1.496\times10^{-6}$ & $-$ & $-8.377 \pm 0.666$ & $-4.576 \times10^{-7} \pm 1.593 \times 10^{-7}$ \\
         
         220 GHz & \phs $ < 8.008 \times10^{-6}$ & $ - $ & $ - $ & $< 6.535 \times10^{-5}$ \\
         \hline
    \end{tabular}
    \label{tab:dipole_fit_data}
\end{table*}

\begin{figure*}
    \centering
    \includegraphics[width=1.9\columnwidth]{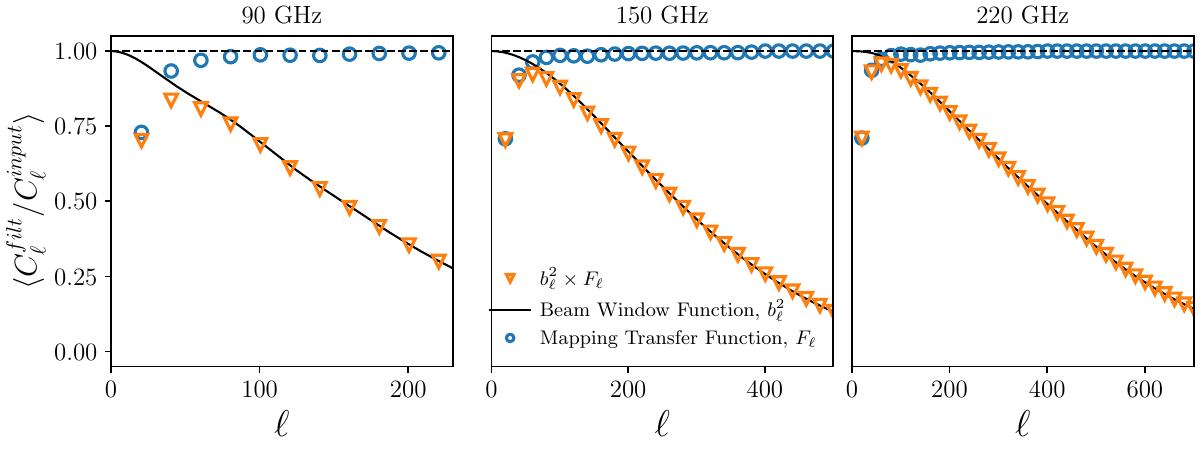}
    \caption{Transfer functions from mapmaking and finite beam angular resolution for CLASS at 90~GHz (left panel), 150~GHz (middle panel), and 220~GHz (right panel). In each panel, the blue points show the mapmaking transfer function. The beam window function is shown as the black curve. The orange points show the combined power spectrum window function, which is the product of the mapmaking transfer function and the beam window function squared. Error bars are not depicted, as they are smaller than the marker size. 
    \label{fig:transfer_func}}
\end{figure*}

To compare to the measured amplitudes, laboratory measurements of oxygen emission lines can be used in conjunction with models of Earth's atmosphere at the CLASS site location and a model of the Earth's magnetic field to simulate the expected atmospheric spectrum. We compare the measured amplitudes with predictions from two modeled atmospheric spectra. The first model is computed following \cite{petroff20}. We fit Equation \ref{eq: dipole} to the simulated signal and the resulting best-fit amplitudes for each frequency are shown as the dark and light blue triangles in Figure~\ref{fig:vatm_spec}. There is significant disagreement between the simulated and measured amplitudes for 90 GHz and 150 GHz. As discussed in Section \ref{subsec:atm systematics}, we cannot explain this discrepancy using known systematic effects in the measurement or the simulation. Based on the error analysis in Section \ref{sec:transfer_func_bias} and Section \ref{subsec:atm systematics}, we conclude that the map amplitudes are reliable and attribute the remaining disagreement between our data and the model to uncertainties in the theory defining the model.

The second model is the Atacama atmospheric model of \citet{spinelli_2011}. The dark and light green squares in Figure \ref{fig:vatm_spec} come from integrating that model over the CLASS bandpasses to predict the observed dipole amplitude. This model agrees well with the CLASS data at 90 GHz and shows better agreement than the \citet{petroff20} model at 150 GHz. Note that this spectrum was calculated for a site 100 meters lower than the CLASS site and at a slightly different latitude and longitude. Additionally, since their model was calculated in 2011, Earth's magnetic field amplitude and direction as observed from the Atacama were different than during the observing period. However, these differences in the simulation inputs do not appear to drive the agreement with the data. Instead, the agreement implies that line shape implementation is the dominant factor. The line-wing shapes are especially relevant for the CLASS bands, and have very different shapes in each model. We attribute the discrepancy in the wings between the two models to different treatments of the line mixing, which is a dominant effect in the lower atmospheric layers.

\subsection{Systematics in the Atmospheric Measurement} \label{subsec:atm systematics}

We consider several sources of systematic uncertainty in the atmospheric modeling and analysis. These are discussed in the following subsections. Importantly, none of these are expected to significantly bias the cosmology maps.

\begin{figure}
    \centering
    \includegraphics[width=\columnwidth]{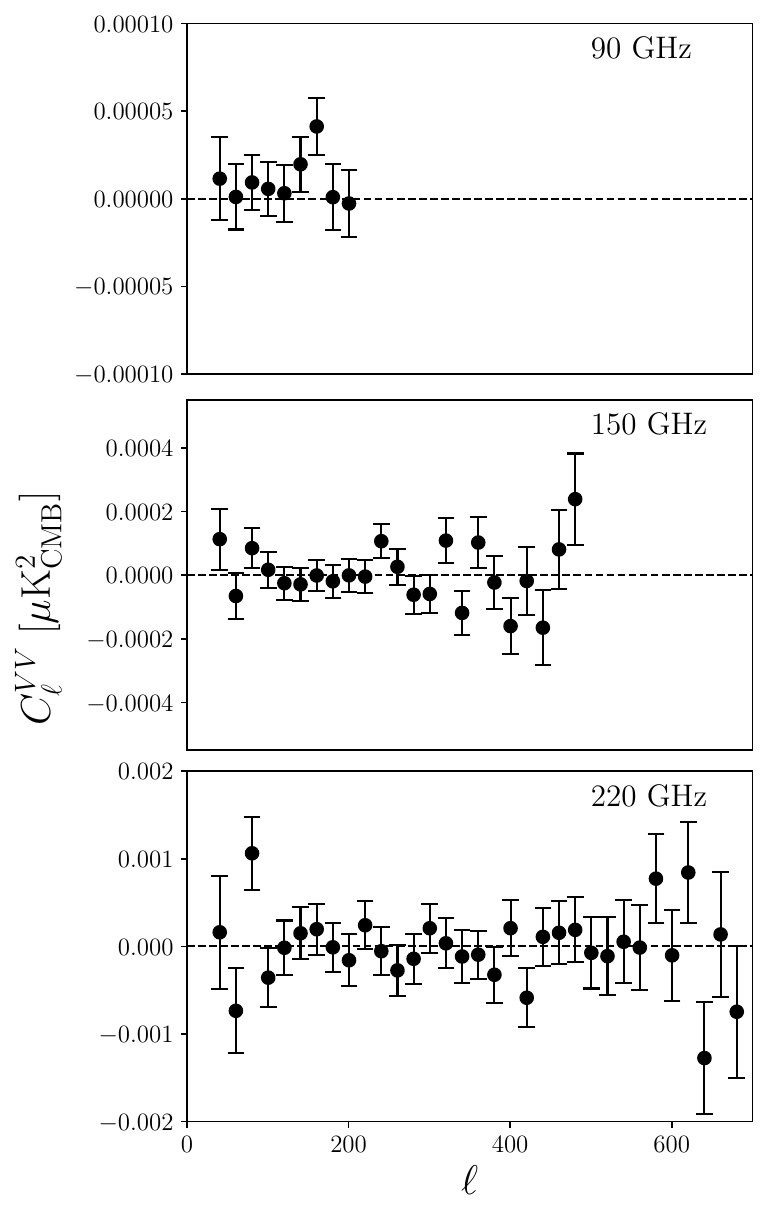}
    \caption{$VV$ power spectra for CLASS 90~GHz (top panel), 150~GHz (middle panel), and 220~GHz (bottom panel). 
    All spectra are consistent with no Celestial circular polarization.
    \label{fig:vv_spectra}}
\end{figure}

\subsubsection{Bandpass Uncertainties} \label{subsubsec:bandpass_unc}

The accuracy of the band center frequency and band edges for the CLASS passbands is a source of uncertainty in the accuracy of the simulation. As shown in Figure~\ref{fig:vatm_spec}, there are nulls in the theoretical spectrum that can change the dipole amplitude depending on the frequencies being integrated over, specifically for the 90~GHz and 150~GHz bands. To account for this, the band center and band edges are sampled from a normal distribution based on the measured bandpasses in \citet{dahal22}, creating a distribution in the simulated amplitudes. The null crossings create asymmetry in these distributions, so the error bars on the blue triangles and green squares in Figure \ref{fig:vatm_spec} include the 16th and 84th percentiles of the simulated amplitude distribution. 

\subsubsection{Impact of Transfer Function Bias} \label{subsubsec:transfer_bias_atm}
From the calculation in Section \ref{sec:transfer_func_bias}, the uncertainty in the modulator transfer function results in a modest systematic error in the derived signal. The impact of this uncertainty on the atmospheric dipole amplitude is included in the error bars on the blue triangles and green squares in Figure \ref{fig:vatm_spec}. 

\subsubsection{Theoretical Uncertainties}

The remaining disagreement between the data and simulated amplitudes from the \citet{petroff20} model is attributed to uncertainties in the shape of the spectral features far in frequency from the line centers. As shown in Figure \ref{fig:vatm_spec}, the theoretical spectrum includes several null crossings, two of which are close to the 90 GHz and 150 GHz bands. Similarly to the uncertainty calculated by shifting the bandpass, slight shifts in the wings around the line centers will create shifts in the amplitudes. From the treatment of potential errors in the map amplitudes (i.e. Sections \ref{sec:transfer_func_bias} and the rest of Section \ref{subsec:atm systematics}), we reasonably trust that our map amplitudes are correct and attribute the remaining disagreement between our data and the model to uncertainties in the theory defining the model. 

\begin{figure}
    \centering
    \includegraphics[width=\columnwidth]{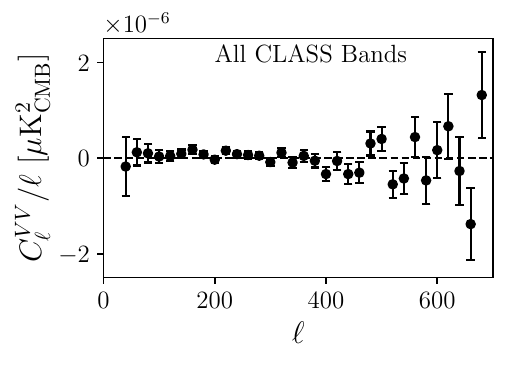}
    \caption{$VV$ power spectrum for all CLASS bands combined. The power spectrum is scaled by $1/\ell$ to more clearly show the scatter of the data points and error bars at all $\ell$. The power spectrum has reduced-$\chi = 0.094$ and reduced-$\chi^2 = 0.681$.
    \label{fig:vv_combo_spectrum}}
\end{figure}

\begin{figure*}[]
    \centering
    \includegraphics[width=0.7\textwidth]{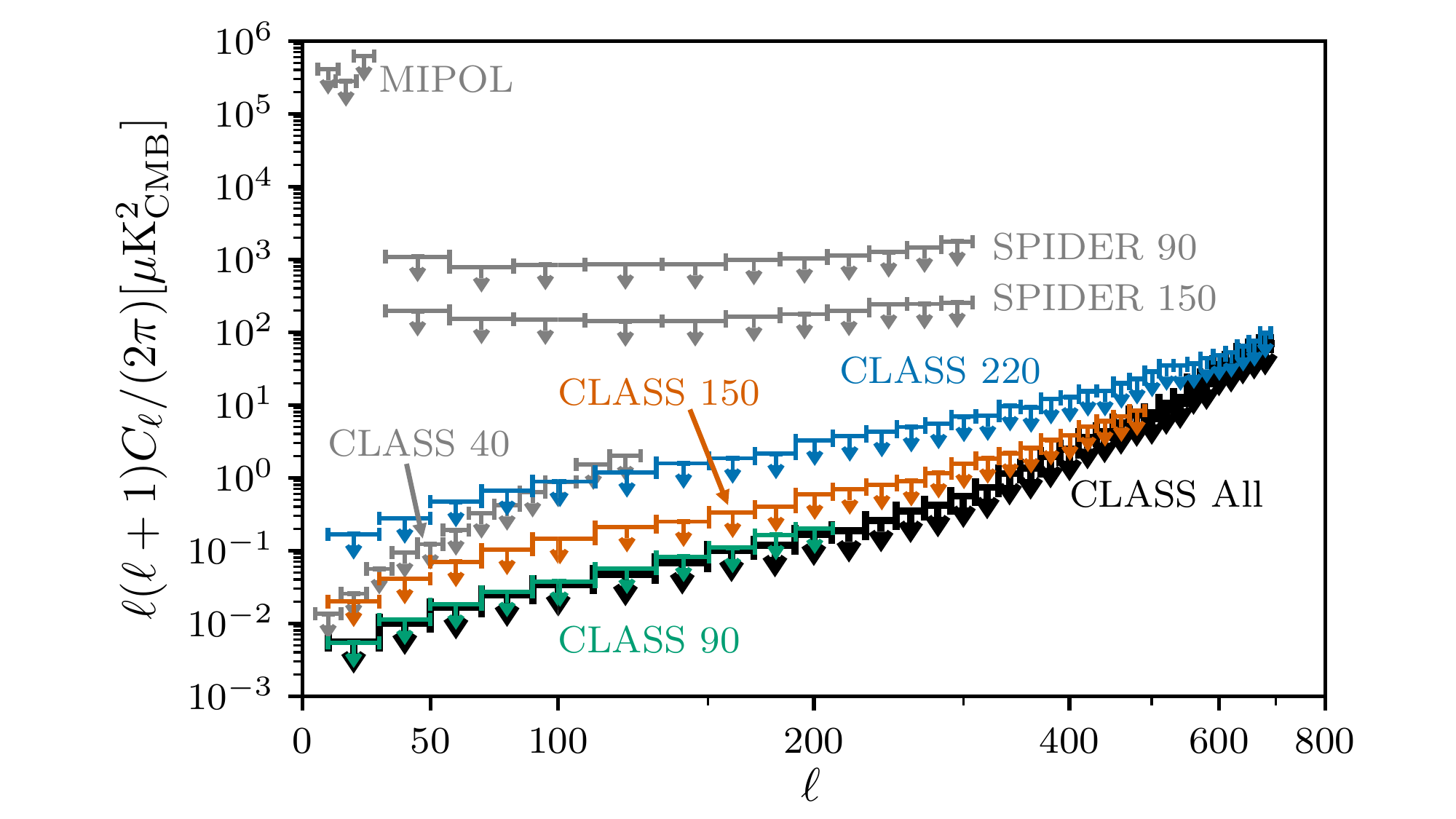}
    \caption{95\% confidence limits on Celestial circular polarization by CLASS at 90, 150, and 220~GHz (this work) compared with prior limits set by CLASS at 40~GHz, as well as limits from the MIPOL and SPIDER experiments. Also plotted is a combined limit from the 90, 150, and 220~GHz bands. 
    \label{fig:vlimits}}
\end{figure*}

\section{Celestial Circular Polarization Constraints}
\label{sec:vv_spectra}
For constraints on the Celestial circular polarization, angular power spectra were computed using the pseudo-$C_\ell$ method of \texttt{PolSpice} \citep{polspice}\footnote{\url{https://www2.iap.fr/users/hivon/software/PolSpice/}} from maps with the atmospheric dipole filtered out. As circular polarization is a scalar, its power spectra are computed in the same way as temperature power spectra, taking into account varying survey weight across the sky and the effect of beam convolution. 
Survey weights were assigned according to the hits maps, which give the number of observations in each pixel and are inversely proportional to the per-pixel variance of the map. Circular polarization cross-spectra were computed between maps from two approximately equal subsets of the data (see Section \ref{sec:mapmaking}) at each frequency (90, 150, and 220~GHz) to avoid instrumental noise bias. Due to visible signal in the Galactic plane in the 90~GHz maps, we mask 10\% of the sky near the Galactic plane using the \textit{Planck} \texttt{f090} mask.\footnote{\url{https://irsa.ipac.caltech.edu/data/Planck/release_2/ancillary-data/previews/HFI_Mask_GalPlane-apo0_2048_R2.00/}}

Simulations are used to evaluate the impact of time-stream filtering on the maps \citep[see][]{Eimer_2023_CLASS_Results, Li23}, through the harmonic domain mapping transfer function:
\begin{equation}
    \hat{C}_b = T_{b} C_{b},  \label{eq:hdtfunc}
\end{equation}

\noindent where $C_b$ refers to the band power of the sky signal. 
The binned transfer function $T_b$ relates these to the angular power spectrum as observed by CLASS, $\hat{C}_b$, ignoring the inter-bin correlation given the bin size of $\Delta \ell=20$. 

The beam window functions are derived from dedicated observations of Jupiter \citep{datta22-spie,datta23}. The mapping transfer functions are characterized using simulations containing only circular-polarization signals (assuming white power spectra), following a procedure similar to that described in \cite{Li23} and \cite{li25}. Figure~\ref{fig:transfer_func} shows the mapping transfer functions and squared beam window functions used in this analysis. 

\begin{figure*}
    \centering
    \includegraphics[width=0.7\textwidth]{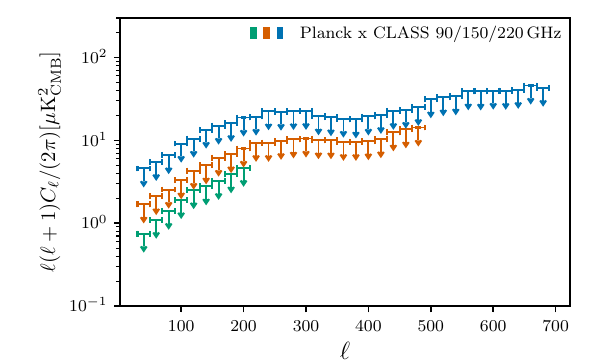}
    \caption{95\% confidence $TV$ upper limits from CLASS (90, 150, and 220~GHz) in cross-correlation with \textit{Planck} PR4 SEVEM maps, applying a Galactic mask that excludes the most contaminated 20\% of the sky. 
    \label{fig:tvlimits}}
\end{figure*}

For the Celestial circular polarization analysis, a set of combined maps was created from the individual 90, 150, and 220~GHz maps. We combine maps in the harmonic domain via a weighted co-add of the data from the three bands assuming that the underlying circular polarization has the same frequency dependence as the CMB. The weights for each band are constructed from the noise covariance matrix estimated using the ensemble of noise simulations described in Section~\ref{sec:sims}, along with the mapping transfer function and beam window functions:

\begin{align}
    a_{\ell m, MV} & = \frac{\sum w_{i,\ell} T^{-1/2}_{i,\ell} b^{-1}_{i,\ell} a_{\ell m, i}}{\sum w_{i,\ell} }, \\
    w_{i,\ell} & = T_{i,\ell} b^2_{i,\ell} N_{i,\ell}^{-1} ,
\end{align}

\noindent where $a_{\ell m}$ are the coefficients in a spherical harmonic expansion of the maps, the summation index $i$ refers to each frequency map (90, 150, and 220 GHz), $T_{i,\ell}$ is the unbinned mapping transfer function, $b_{i,\ell}$ is the beam window function, and $N_{i,\ell}^{-1}$ are the diagonal elements of the inverse of the noise covariance matrix. As the mapping transfer function is defined for the power spectrum, $C_\ell \propto a_{\ell m}^2$, we take a square root of the transfer function when combining $a_{\ell m}$. Weights are constructed for each $\ell$ independent of $m$.

 \begin{figure}
    \centering
    \includegraphics[width=\columnwidth]{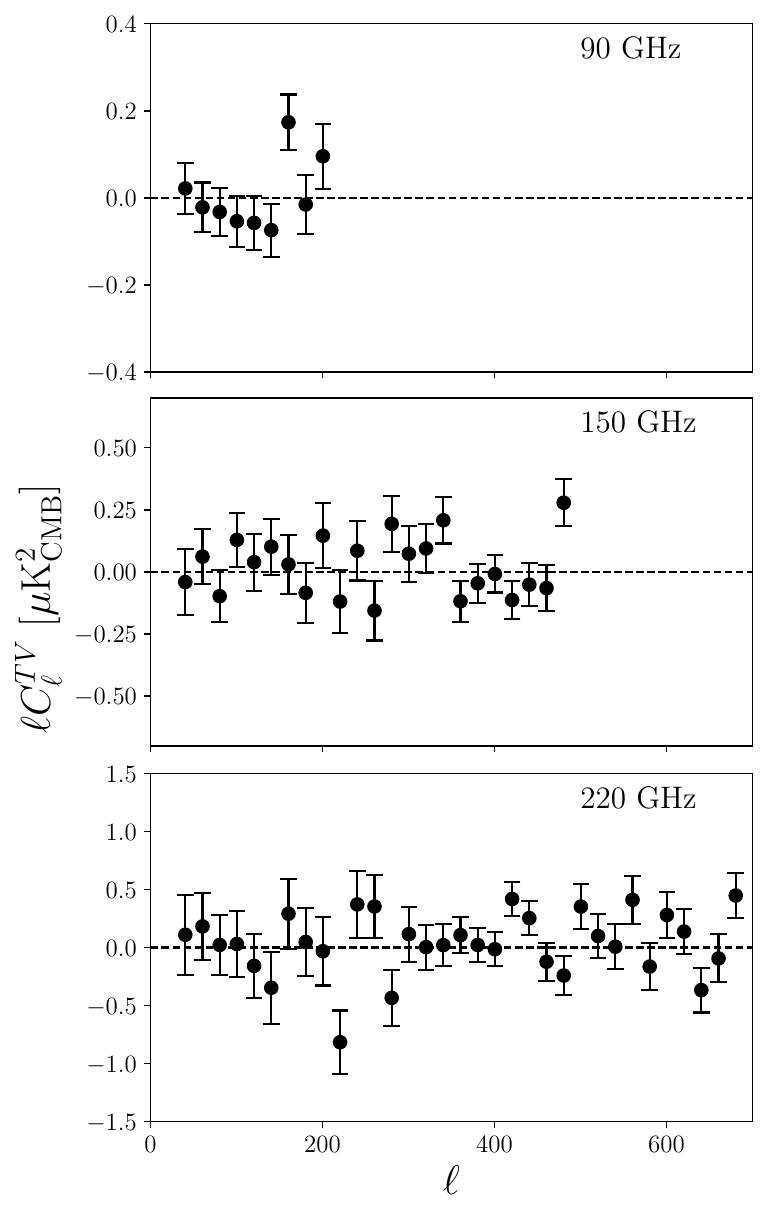}
    \caption{$TV$ power spectra for \textit{Planck} SEVEM temperature and CLASS. 
    The spectra are consistent with no $TV$ signal. The power spectra are scaled  by $\ell$ to more clearly show the scatter of the data points and error bars at high $\ell$. Note that the $y$ axis range changes with frequency band.
    \label{fig:tv_spectra}}
\end{figure}

As with the individual frequency maps, combined maps were created for two subsets of the data of approximately equal weight for use in power spectrum estimation through cross-spectra to avoid noise bias. 
The CLASS 40~GHz maps were not included in the map combination, as they would not contribute significantly to the map depth at any $\ell$.

The resultant $VV$ power spectra from CLASS data at 90, 150, and 220~GHz are shown in Figure~\ref{fig:vv_spectra}. These spectra are consistent with no Celestial circular polarization with reduced-$\chi^2 =$ (0.972, 1.155, 0.859) for (10, 24, 34) degrees of freedom with PTE = (0.449, 0.246, 0.655) for the (90, 150, 220)~GHz bands, respectively. The reduced-$\chi$ (sum of deviations) = (0.599, $0.023$, and $-0.029$) with PTE = (0.030, 0.421, 0.550). The errors on the binned power spectra are estimated using 500 cross-spectra of pairs of maps pulled from 100 simulations of map splits. 
The combined power spectrum from all frequencies is shown in Figure~\ref{fig:vv_combo_spectrum}. This spectrum is consistent with no signal with reduced-$\chi = 0.094$ and reduced-$\chi^2 = 0.681$.

The distribution of bin powers from noise simulations is used to estimate upper limits (95\% confidence) on Celestial circular polarization in the CLASS frequency bands versus multipole moment as shown in Figure~\ref{fig:vlimits}, along with limits from prior CLASS work at 40~GHz~\citep{Eimer_2023_CLASS_Results} and the MIPOL and SPIDER instruments~\citep{MIPOL_V_Limit2013, SPIDER_V_Limit_2017}. These results extend the range of multipoles constrained by CLASS to $\ell=(200, 500, 700)$ at $(90, 150, 220)$~GHz, respectively. The limits set by CLASS at 90~GHz improve upon prior limits set by CLASS at 40~GHz out to $\ell=125$ by factors of 5--25, depending on angular scale. 

\section{Temperature--Circular--Polarization Cross Spectrum}
\label{sec:tv_spectra}

In this work we constrain the $TV$ cross-spectrum using \textit{Planck} temperature maps and CLASS circular polarization maps at 90, 150, and 220~GHz. The \textit{PR4} CMB maps made with the SEVEM algorithm by the \textit{Planck} team provide the CMB temperature anisotropy estimates required to compute $TV$~\citep{planck18IV, 2006MNRAS.370.2047L, 2016MNRAS.459..441F}.

A total of 600 \textit{Planck} simulations are provided, separately, for the CMB component estimate and for the noise. 
The CMB simulations include the sample variance from the CMB and foreground residuals from component separation.
CLASS noise is estimated using simulations as described in Section~\ref{sec:sims}.
The \textit{Planck} SEVEM maps have an equivalent beam size of $5^\prime$, which is corrected for in the $TV$ spectra computed below.

\subsection{$TV$ Cross-Spectrum Results}

The $TV$ cross-spectrum is computed as described for $VV$ above with the \textit{Planck} \texttt{f080} mask that removes 20\% of the sky most heavily contaminated by the Galaxy.\footnote{\url{https://irsa.ipac.caltech.edu/data/Planck/release_2/ancillary-data/previews/HFI_Mask_GalPlane-apo0_2048_R2.00/}} The uncertainties on the cross power spectrum $TV$ are derived from 500 independent cross-spectra of pairs of simulations from \textit{Planck} (signal and noise) and CLASS (noise) for signal--noise and noise--noise spectra separately. Given that the \textit{Planck} maps are temperature signal dominated, the signal--noise cross-spectra dominate the upper limits for $TV$ from \textit{Planck} $\times$ CLASS.

The 95\% confidence level upper limits on $TV$ are shown in Figure~\ref{fig:tvlimits}. The $TV$ spectra are shown in Figure~\ref{fig:tv_spectra}. These spectra 
are consistent with zero with reduced-$\chi^2 = $ (1.450, 1.398, 1.624) for (10, 24, 24) degrees of freedom with PTE = (0.174, 0.106, 0.016) for the $(90, 150, 220)$~GHz bands, respectively. The reduced-$\chi$ (sum of deviations) = (0.007, 0.196, 0.232) with PTE = (0.464, 0.152, 0.092). 
These are the first constraints placed on $TV$ cross-correlation in the CMB to date. 

\section{Conclusion}
In this paper, we present first results from the CLASS experiment on atmospheric and Celestial circular polarization at frequencies of 90, 150, and 220~GHz, building on prior results at 40~GHz~\citep{Padilla_2020_CLASS_Circular, petroff20, Eimer_2023_CLASS_Results}. We measure the atmospheric circular polarization from Zeeman splitting of oxygen emission lines, which produces a dipole-like pattern in horizontal coordinates aligned with the Earth's magnetic field direction. The dipole signal is detected at 40 GHz, 90 GHz, and 150 GHz. 

We find that the data prefer the \citet{spinelli_2011} atmospheric model over the \citet{petroff20} model at 90 and 150 GHz. The disagreement in those bands is attributed to differing treatment and data for the line mixing parameters between the models, which introduce additional complexity in the line wings. The line-wing shapes are especially relevant for the CLASS bands, and have very different shapes in each model, including different locations of null crossings. 

We find no evidence for Celestial circular polarization and present improved constraints on the circular polarization power spectrum ($VV$). CLASS places an upper limit of $4.5 \times 10^{-3}$~$\mu$K$_{\text{CMB}}^2$ on $VV$ at 95\% confidence at large angular scale ($10 < \ell < 30$). This work extends prior limits on $VV$ from CLASS at 40~GHz (covering $\ell <$ 120) to a broader range of angular scales ($\ell <$ 200, 500, and 700 for 90, 150, and 220~GHz, respectively). The improved sensitivity of the CLASS 90~GHz band in particular tightens upper limits on Celestial circular polarization through the $VV$ power spectrum by factors of 5--25 compared with prior CLASS 40 GHz only limits, depending on angular scale. 

Finally, we place the first constraints on the cross-correlation between CMB temperature and circular polarization ($TV$) through combination with \textit{Planck} SEVEM maps, finding it to be consistent with zero. CLASS places an upper limit of $5.8 \times 10^{-1}$~$\mu$K$_{\text{CMB}}^2$ on $TV$ at 95\% confidence at large angular scale ($30 > \ell > 50$).

\begin{acknowledgments}
We thank Massimo Gervasi, Sebastiano Spinelli, Giulio Fabbian, Andrea Tartari, and Mario Zannoni for kindly sharing their model atmospheric spectrum. We acknowledge the Astronomical Sciences Section of the National Science Foundation Directorate for Mathematical and Physical Sciences (MPS) 
for their support of CLASS under Grant Numbers 0959349, 1429236, 1636634, 1654494, 2034400, 2109311, and 2442928. We thank Johns Hopkins University President R. Daniels and Dean C. Celenza for their steadfast support of CLASS. We further acknowledge the very generous support of Jim Murren and Heather Miller (JHU A\&S '88), Matthew Polk (JHU A\&S Physics BS '71), David Nicholson, and Michael Bloomberg (JHU Engineering '64).  The CLASS project employs detector technology developed in collaboration between JHU and Goddard Space Flight Center under several previous NASA grants. Detector development work at JHU was funded by NASA cooperative agreement 80NSSC19M0005. 
Part of this work was carried out at the Advanced Research Computing at Hopkins (ARCH) core facility (rockfish.jhu.edu), which is supported by the National Science Foundation (NSF) grant number OAC1920103. CLASS is located in the Parque Astron\'omico Atacama in northern Chile under the auspices of the Agencia Nacional de Investigaci\'on y Desarrollo (ANID).

This paper uses data products derived from observations obtained with \textit{Planck} (http://www.esa.int/Planck), an ESA science mission with instruments and contributions directly funded by ESA Member States, NASA, and Canada.

The authors declare the use of Claude Sonnet 5.0, a form of Generative Artificial Intelligence (GAI), to refine visualizations of results and for proofreading of the manuscript. All tasks were performed under full human supervision.


\software{
numpy \citep{numpy20}, 
scipy \citep{scipy}, 
matplotlib \citep{matplotlib},
astropy \citep{astropy}, 
HEALPix \citep{healpix},
camb \citep{camb},
pysm \citep{pysm},
PolSpice \citep{polspice},
NaMaster \citep{Alonso:2018jzx}
}
\end{acknowledgments}

\newpage
\bibliographystyle{aasjournalv7mod}
\bibliography{class_vpol, class_pub, cmb}

\end{document}